\documentclass[11pt]{article}
\usepackage{ifthen}
\usepackage{fullpage}

 \usepackage{graphicx}

\usepackage{amsfonts}
\usepackage{wrapfig}
\usepackage[font=footnotesize,labelfont=bf]{caption}

\newcommand{\newparentheses}[3]{%
  \expandafter\newcommand\csname #1\endcsname[1]{#2##1#3}%
  \expandafter\newcommand\csname #1L\endcsname[1]{\bigl#2##1\bigr#3}%
  \expandafter\newcommand\csname #1XL\endcsname[1]{\Bigl#2##1\Bigr#3}%
  \expandafter\newcommand\csname #1V\endcsname[1]{\left#2##1\right#3}}

 \newparentheses{parens}{(}{)}

\makeatletter
\newcommand{\onenewattribute}[4]{%
  \@ifundefined{#2}{\let\@@def\newcommand}{\let\@@def\renewcommand}%
  \expandafter\@@def\csname #2\endcsname[1][]{%
    \def\first@arg{##1}\csname @#2\endcsname}%
  \@ifundefined{@#2}{\let\@@def\newcommand}{\let\@@def\renewcommand}%
  \expandafter\@@def\csname @#2\endcsname[2][]{%
    \ifthenelse{\equal{#1}{sub}}%
    {\csname @@#2\endcsname{##1}{\first@arg}{##2}}%
    {\csname @@#2\endcsname{\first@arg}{##1}{##2}}}
  \@ifundefined{@@#2}{\let\@@def\newcommand}{\let\@@def\renewcommand}%
  \expandafter\@@def\csname @@#2\endcsname[3]{%
    \ifthenelse{\equal{##1}{}}%
    {\ifthenelse{\equal{##2}{}}%
      {#3\csname #4\endcsname{##3}}%
      {#3_{##2}\csname #4\endcsname{##3}}}%
    {\ifthenelse{\equal{##2}{}}%
      {#3^{##1}\csname #4\endcsname{##3}}%
      {#3_{##2}^{##1}\csname #4\endcsname{##3}}}}}
\newcommand{\newattribute}[3][sub]{%
  \onenewattribute{#1}{#2}{#3}{parens}%
  \onenewattribute{#1}{#2L}{#3}{parensL}%
  \onenewattribute{#1}{#2XL}{#3}{parensXL}%
  \onenewattribute{#1}{#2V}{#3}{parensV}}
\makeatother

\newattribute{OhOf}{\mathrm{O}}
\newattribute{ThetaOf}{\Theta}
\newattribute{OmegaOf}{\Omega}
\newattribute{ohOf}{\mathrm{o}}
\newattribute{omegaOf}{\omega}

\newattribute{sort}{\mathrm{sort}}
\newattribute{psort}{\mathrm{sort}_P}
\newattribute{onesort}{\mathrm{sort}_1}
\newattribute{scan}{\mathrm{scan}}
\newattribute{pscan}{\mathrm{scan}_P}

\newtheorem{definition}{Definition}

\newcommand{\reals}{\mathbb{R}}

\begin{document}

\title{Empirical Evaluation of the Parallel Distribution Sweeping
  Framework on Multicore Architectures}

\author{Deepak Ajwani \\
Bell Laboratories Ireland\\
Dublin, Ireland\\
Email: deepak.ajwani@alcatel-lucent.com\\
\and
Nodari Sitchinava\\
Institute of Theoretical Informatics\\
Karlsruhe Institute of Technology\\
Karlsruhe, Germany\\
Email: nodari@ira.uka.de\\
}

\date{}
\maketitle
\begin{abstract}
In this paper, we perform an empirical evaluation of the Parallel External
Memory (PEM) model in the context of geometric problems. In particular, we
implement the parallel distribution sweeping framework of Ajwani, Sitchinava and Zeh to solve batched 1-dimensional stabbing max 
problem. While modern processors consist of sophisticated memory
systems (multiple levels of caches, set associativity, TLB,
prefetching), we empirically show that algorithms designed in simple
models, that focus on minimizing the I/O transfers between shared
memory and single level cache, can lead to efficient software on
current multicore architectures. Our implementation exhibits
significantly fewer accesses to slow DRAM and, therefore,
outperforms 
traditional approaches based on plane sweep and two-way divide and
conquer. 
\end{abstract}

\section{Introduction}
Modern multicore architectures have complex memory systems
involving multiple levels of private and/or shared caches, set
associativity, TLBs, and prefetching effects. It is considered
challenging to design and even engineer algorithms to directly
optimize the running time on such architectures~\cite{kang:multicore-challenge}.
Furthermore, algorithms optimized for one
architecture may not be optimal for another. To address these issues, various
computational models~\cite{arge:pem-sorting,blelloch:multicore-dnc,blelloc:pco,chowdhury:multicore-gauss-elim,chowdhury:multicore-dyn-prog}
have been proposed in recent years. These computational models are
simple (usually assuming only two levels of memory hierarchy, out of
which one is shared) as they abstract away the messy architectural
details. Also, the performance metric of these models involve a single
objective function such as minimizing shared memory accesses. The
simplicity of these models allows the design of practical algorithms that are
expected to work well on various multicore
architectures. It also allows us to compare the relative
performance of algorithms theoretically.

The success of a computational model crucially depends on how well the
theoretical prediction of an algorithm in that model matches the actual running time on real systems. 
Unfortunately so far, there has been little empirical work (such as~\cite{tang:stencil}) to
evaluate the predictions of algorithmic performance using these
models on real multicore architectures. It is not even clear if
these models can lead to the design of algorithms that are faster on
current multicore systems (with 2 - 48 cores) than
those designed in the traditional RAM model, external memory model and the PRAM model. In fact,
many of the algorithms 
designed in these models for multicores seem quite sophisticated and
are likely to have high constant factors that can pay off only for
architectures with hundreds of cores. This state of affairs 
is in sharp contrast with the sequential cache-efficient models,
where a considerable empirical work (e.g.,~\cite{bender:streaming_b_trees,brodal:co_sorting})
evaluating the algorithms on real systems exists.

At the core of the debate for the computational model is the choice of
the performance metric that an algorithm designer should optimize for
the current multicore systems. In the
traditional RAM (and PRAM) model of computation, the algorithms are designed to
minimize the number of instructions (and parallel instructions) executed by the algorithm. 
The {\em external memory (EM)} 
model~\cite{aggarwal:io-model} when applied to cached memories (e.g., see~\cite{mehlhorn:associativity}) aims at minimizing the cache
misses, ignoring the number of instructions. The {\em parallel external
memory (PEM)} model~\cite{arge:pem-sorting} aims at minimizing the number of {\em parallel} cache
misses. 

In this work, we demonstrate that algorithms designed in simple models, that
focus on minimizing the parallel I/O transfers between shared memory and a
single level cache, can lead to a software performing great in practice on real
multicore systems.
 For this purpose, we consider the algorithms to solve the problem of answering
batched planar orthogonal stabbing-max queries.  This problem is a fundamental
geometric primitive and together with its variants is used as subroutines in
solutions of many popular geometric problems such as point location in an
orthogonal subdivision of the plane, orthogonal ray shooting, batched (offline)
dynamic predecessor queries in 1-dimensional array and batched union-find.
Also, this problem has been well-studied in various computational models and
many different optimal solutions for it are known in these models. Thus, it
provides a test-bed for evaluating the efficacy of theoretical analysis in
various models on real multicore architectures. Another reason for selecting
this non-HPC application is that the ratio of memory accesses to computation in
the solutions of this problem is similar to that of many data-intensive
geometric applications. For instance, our engineered PEM solution for this
problem is based on the parallel distribution sweeping framework and this
framework has been used for designing a wide range of other geometric
algorithms in the PEM model~\cite{ajwani:pem-geom,ajwani:ipdps11} and a basis
for PEM data structures~\cite{sitchinava:pem-buffer-tree}.

We empirically compare the different
solutions and show that a carefully engineered solution based on an algorithm in the
PEM model gives the best performance on various 
 multicore systems, outperforming traditional approaches based on plane sweep, sequential distribution
sweeping and two-way divide-and-conquer. Using hardware profilers, we
show that this solution exhibits significantly fewer number of
accesses to slow DRAM which is correlated with the improved running time. 

Since the cache line on modern systems is typically 64 bytes,
I/O-efficient solutions also need to be work-efficient to
compete with RAM algorithms. In other words, the total number of
instructions of a cache-efficient algorithm should asymptotically match that of the
best RAM solution. Therefore, we design an algorithm that is both I/O-optimal and work-efficient. To the best of our knowledge, this is the first
work-efficient I/O-optimal algorithm for this problem.

\section{Computational Models}

\label{sec:model}
\textbf{External Memory Model.} The widely used external memory  model or the I/O model by Aggarwal
and Vitter~\cite{aggarwal:io-model} assumes a two level memory
hierarchy. The internal memory has a limited size and can hold at most
$M$ objects (points/line-segments) and the external memory has a
conceptually unlimited size. The computation can only use the data in
the internal memory, while the input and the output are stored in the
external memory. The data transfer between the two memories happens in
blocks of $B$ objects. The measure of
performance of an algorithm is the number of I/Os (cache misses) it performs.  The
number of I/Os needed to read $n$ contiguous items from the external memory is
$\scan{n} = \Theta(n/B)$.  The number of I/Os required to
sort $n$ items is $\sort{n}=\Theta((n/B)\log_{M/B}(n/B))$. 
For all realistic values of $n$, $B$, and $M$, $\scan{n}<\sort{n} < 
n \log_2{n}$.

\begin{wrapfigure}{l}{2.2in} {\centering \includegraphics{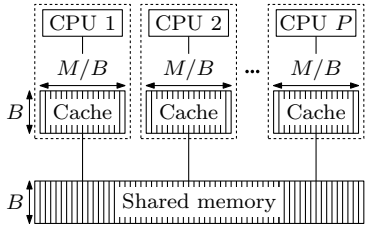} \caption{The PEM
model} \label{fig:pem-model} } \end{wrapfigure}

\textbf{Parallel External Memory (PEM) Model.} 
The \emph{parallel external memory} (PEM) model~\cite{arge:pem-sorting}
is a simple parallelization of the EM model. It  consists of $P$ processors, each with a \emph{private cache} of
size~$M$ (see Figure~\ref{fig:pem-model}).  Processors communicate with each
other through access to a \emph{shared memory} of conceptually unlimited size.
Each processor can use only data in its private cache for computation.
%
The caches and the shared memory are divided into \emph{blocks} of size $B$.  Data
is transferred between the caches and shared memory using \emph{parallel
input-output} (I/O) operations.  During each such operation, each processor can
transfer one block between shared memory and its private cache.  The cost of an
algorithm is the number of I/Os it performs.  Concurrent reading of the same
block by multiple processors is allowed but concurrent block writes are
disallowed (similar to a CREW PRAM).  The cost of sorting in this model is
$\psort{n} = \OhOfL{\frac{n}{PB}\log_{M/B}\frac{n}{B}}$ parallel
I/Os, provided $P \le n / B^2$ and $M = B^{\OhOf{1}}$~\cite{arge:pem-sorting}.

The PEM model provides the simplest possible abstraction of current
multicore chips, focusing on the fundamental I/O issues that need to
be addressed when designing algorithms for these architectures,
similar to the I/O model \cite{aggarwal:io-model} in the sequential
setting. 
\section{1-D Stabbing Max Algorithms}
In this section, we describe various algorithms that we
implemented and used for our experimental study. We begin with
formally describing the problem.

\begin{definition}[Batched 1-D Stabbing-Max Problem]
Given a set of $n$ horizontal line segments and points on the plane, report for
each point the closest segment that lies directly below it.
\end{definition}

\textbf{RAM algorithm.} 
In the classical RAM model, this problem is solved using the sweep line
paradigm~\cite{shamos:line-sweep,bentley:line-sweep-intersections}.  We sweep a
hypothetical vertical line across the plane in increasing $x$-coordinate and
perform some computation at each segment endpoint or query point. We maintain
an ordered set $A$ of {\em active segments} --- all segments which intersect
the sweep line, ordered by the $y$-coordinates. A segment is inserted into $A$
when the sweep line encounters its left endpoint and removed when it encounters
the right endpoint. An answer to a query point $q$ is the segment in $A$ with
the largest $y$-coordinate that is smaller than the $y$-coordinate of $q$,
i.e., the predecessor of $q$ in $A$ according to the $y$-ordering.

For $n$ line segments and query points, there are $\OhOf{n}$ insertions,
deletions and predecessor searches in $A$.  Since each of these operations can
be performed in $\OhOf{\log{n}}$ time by maintaining $A$ as a balanced binary
search tree, the total complexity of this algorithm is $\OhOf{n\log{n}}$
instructions.

\textbf{Sequential I/O-optimal solution.}
The sequential I/O-efficient solution for this problem proceeds using the
distribution sweeping framework of Goodrich et
al.~\cite{goodrich:external-geometry} as follows. 

Let $r_q$ be a variable associated with each query point $q$ which we will use to
store the answer. Initially $r_q$ is initialized to 
a virtual horizontal line $y = -\infty$.

We partition the space into $K = \min\{M/B, n/M\}$ vertical slabs $\sigma_1,
\dots, \sigma_K$, so that each slab contains equal number of points (endpoints
of horizontal segments or query points) and perform a sweep of the input by
increasing $y$-coordinate. During the sweep we maintain for each slab
$\sigma_i$ a segment $s_{\sigma_i}$ which is the highest segment that spans
$\sigma_i$ encountered by the sweep. When the sweep line encounters the query
point $q \in \sigma_i$, we update $r_q$ with $s_{\sigma_i}$ iff
$y(s_{\sigma_i}) > y(r_q)$. During the sweep we also generate slab lists
$Y_{\sigma_i}$. A copy of a query $q$ (resp., segment $s$) is added to
$Y_{\sigma_i}$ if $q$ (resp., at least one of the endpoints of $s$) lies in
slab $\sigma_i$. The sweep is followed by a recursive processing of each slab,
using $Y_{\sigma_i}$ as input for the recursive call.  The recursion terminates
when each slab contains $\OhOf{M}$ points and the problem can be solved in
internal memory, for example, by using the plane sweep algorithm.

Note, that if the initial objects are sorted by $y$-coordinates, we can
generate the inputs $Y_{\sigma_i}$ for the recursive calls sorted by
$y$-coordinate during the sweep. Thus, the sweep at each of $\OhOf{1 + \log_K
(n/M)}$ recursive levels takes $\OhOf{n/B}$ I/Os and the total I/O
complexity of distribution sweeping is $\OhOfV{\frac{n}{B}(1+\log_K n/M)} =
\sort{n}$ I/Os.

\textbf{Work-optimal solution.} 
Note that a naive implementation of the sweep in
internal memory might potentially result in updating $K$ different variables
$s_{\sigma_i}$ whenever a segment is encountered during the sweep. This could
lead to $\OhOf{K n}$ instructions at each recursive level, resulting in total
$\OhOf{K n \log_K n}$ instructions, which is larger than $\OhOf{n \log_2 n}$
instructions of the plane sweep algorithm. At the same time, the plane sweep
algorithm could result in up to $\OhOf{n \log_2 n}$ I/Os, which is larger than
$\sort{n}$ I/Os of the above algorithm. 

To achieve optimal internal computation time while maintaining the optimal
$\sort{n}$ I/O complexity we store segments $s_{\sigma_i}$ in a segment tree
$T$ over $K$ intervals defined by the slabs $\sigma_i$. Since, we are
interested only in segments that fully span the slabs, each segment is stored
only in one node.  Also, at each node we store only the highest segment
encountered up to that point in the sweep. Thus, $|T| = \OhOf{K}$, i.e. $T$
fits in internal memory.  Consider the nodes on the root to leaf path which
correspond to the intervals containing $q$. We update $r_q$ to the highest
segment stored at these nodes. Thus, maintaining $T$ and updating $r_q$ takes
$\OhOf{\log_2 K}$ instructions per update/query, and over $\OhOf{1+\log_K N/M}$
recursive levels of distribution sweeping adds up to at most $\OhOf{n \log_2
n}$ instructions, which is optimal.

\textbf{Parallel External Memory Solution.}
The PEM solution is based on the {\em parallel} distribution sweeping framework
introduced by Ajwani et al.~\cite{ajwani:pem-geom}. It differs from the
sequential distribution sweeping by recursively dividing the plane into $K :=
\max\{2, \min\{\sqrt{n/P}, {M/B}, P\}\}$ vertical slabs\footnote{The
explanation for this choice of $K$ can be found in~\cite{arge:pem-sorting}.}
and performing the sweep in parallel using all $P$ processors. During
recursion, the slabs are processed concurrently using sets of $\ThetaOf{P/K}$
distinct processors per slab. The parallel recursion proceeds for $\OhOf{\log_K
P}$ rounds, until there are $\ThetaOf{P}$ slabs remaining, at which point, each
slab is processed concurrently using a single processor running the sequential
I/O-efficient solution.

To perform the sweep of a single recursive level in parallel using multiple
processors, each processor performs distribution sweeping on an equal fraction
of the input. Note, that such a sweep sets the values of $r_q$ correctly only
if both the query $q$ and the spanning segment $s_{\sigma_i}$ below it are
processed by the same processor. To correct the values $r_q$ across the
boundaries of the parallel sweeps we perform a round of parallel reduction on
segments and queries using MAX associative operator~\cite{blelloch:scan}.
Finally, we compact the portions of slab lists $Y_{\sigma_i}$ generated by
different processors into contiguous slab lists to be used as input for
recursive calls. The details of the algorithm follow directly
from~\cite{ajwani:pem-geom} but are also presented in
Appendix~\ref{app:pem-solution} for completeness. 

The parallel I/O complexity of the above algorithm is $\OhOf{\psort{n}}$ I/Os.

\textbf{Work-optimal solution.} Similar to the sequential I/O model, we can
achieve work optimality in the PEM model algorithm by maintaining a segment tree $T$ on
the $K$ child slabs. In this case, all processors keep their
own copy of $T$ and the parallel reduction (using MAX operator) is
performed over not only the $K$ leaves, but also the $K-1$ internal nodes of $T$. This
does not affect the asymptotic number of parallel I/Os, but makes the scheme
work-optimal, i.e. $\OhOf{\frac{n}{P}\log n}$ instructions per
processor.

\textbf{2-way Distribution Sweeping.} As a PRAM solution, we consider a recursive
2-way distribution sweeping algorithm. This framework is akin to
divide-and-conquer paradigm, that is archetype for many PRAM
algorithms. The 2-way distribution is continued recursively till the slab size is smaller
than a fixed constant and at that stage, plane sweep algorithm is used
as a base case. The distribution step is a simplified version
of the corresponding step in the PEM algorithm, as the considerations
of work-optimality no longer apply.

\section{Implementation Details}
We implemented our algorithms in C++, using OpenMP for
parallelization. The engineered implementation uses some simple
techniques to improve the running time of the theoretical algorithm,
while trying to preserve its worst-case asymptotic guarantee on the
number of shared cache accesses.

The parallel distribution sweeping calls for setting the branching parameter at
$K = \max\{ 2,$ $\min \{M/B, \sqrt{n/P}, P\}\}$. The parameter $M$ also defines the size of
the recursive base case.  We experimentally determine the best choice of $M$.
In particular we found that setting $M$ to be a large fraction (e.g., 1/3 or
1/4) of the L3 cache results in best running times. 

Having determined $M$, we observe that for computing $K$, in our compute
systems the number of processors (up to 12) is far below the other two terms.
Thus, the first recursive level is always a single $P$-way parallel
distribution sweeping round, which results in $P$ vertical slabs each of which
can be processed independently of others in the consequent phases.  Thus, after
the parallel distribution, each of $P$ resulting vertical slabs is assigned to
a separate thread which processes it using a sequential distribution
sweeping algorithm. 

To perform the parallel sweep, we divide the input based on the
$y$-coordinate among the $P$ threads, conceptually, assigning a horizontal slab
of objects to each thread.  The thread with the smaller ID gets the lower $y$
values.  This can be viewed as a $P \times P$ matrix where the columns
correspond to the different slabs and the rows correspond to the different
threads. 

We perform the prefix sum on the $P \times P$ array sequentially as the
overheads associated with the synchronization barrier of OpenMP are too high to
justify this operation in parallel.\footnote{In our experiments, performing this step
sequentially takes less than a millisecond, while the overall running time is
in dozens or hundreds of seconds.}

We combine the second scan of the data (due to reduction) with the step of
compacting child slab lists into contiguous vectors.  During the compaction,
each processor $p_j$ copies all partial chunks of child slab $\sigma_j$ into
the contiguous space. Note, the propagation of the results of the prefix sums
simply needs to update the result of each query point that had been assigned
the sentinel line $y = -\infty$ with the result of the prefix sums value.
Thus, the propagation of the prefix sums values can be performed during  this
copying process. 

Next, we process the $P$ child slabs in parallel using sequential
distribution sweeping. This recursively
subdivides the slabs till the pre-specified threshold $M$ is reached. When
generating the input lists for the child slabs, we also store the
total number of segments and query points for the child slabs. If for
any slab, either the number of segments or query
points is zero, we do not process it or its child invocations any further.

\textbf{Space efficiency.}
We carefully engineered our algorithms to reduce the space
requirement of our implementations considerably. This is done while ensuring that
the running time of our implementations is not affected by the space
reduction. We provide more details of this in Appendix~\ref{sec:space}.

\textbf{Randomized vs. deterministic computation of slab boundaries.} 
Deterministic identification of slab boundaries such that all the child slabs
at each level of recursion contain the same number of objects, requires sorting
the input based on the $x$-coordinate and storing $\OhOf{n/M}$ equally spaced
entries of the sorted input in a separate array. We avoid the extra sort by
instead determining the slab boundaries by partitioning the space into uniform
vertical slabs.  This optimization works well for random input, but in the
worst case can result in the recursion depth as large as
$\OhOf{\log_K{\delta}}$, where $\delta$ is the {\em spread} of the point set --
the ratio between the largest and the smallest (horizontal) distance between a
pair of points.  In case of a large base case of the recursion and randomized
input, this is not an issue. But in the case of double precision coordinates,
the worse case analysis dictates that the depth of the recursion can be very
large. 

\textbf{Constant factors vs. EM implementation.}
The I/O complexity of the sequential distribution sweeping framework is
$\OhOf{n/B (1+\log_{K} n/M)}$, where $K = \min\{M/B, n/M\}$. Since in our
experimental settings $K = n/M$, there are only $2$ recursive levels: one for
distribution sweeping and one for the sweep line at the base case. Thus, the
implementation performs two sequential scans of the input. 

In the parallel version, we have to perform two additional scans. Specifically,
we perform one extra recursive step -- the parallel distribution. During this
step, each processor scans $n/P$ items and writes them out into its private
child slabs. After the prefix sums, which takes negligible amount of time, we
must (a) propagate the result of the prefix sums to the queries that contain
only sentinel values as the result and (b) construct each child slab in
contiguous space. As described earlier, we combine these two tasks into a
single scan. 

Thus, combined with the two scans of the parallel recursive invocation of the
sequential distribution sweeping, the parallel implementation performs a total
of four scans of the input, i.e., twice as many as the sequential version. Since
all scans are performed in parallel and in expectation each child slab contains
equal number of items, the total I/Os performed by each processor is $2/P$
times the number of sequential I/Os, and (ignoring the speedup due to faster
parallel internal computation) we should expect the speed up of $P/2$ on $P$
processors.

\textbf{Sorting.} To perform the initial sorting of the input by the $y$-coordinate, we used the
sorting implementation from the C++ Multicore Standard Template
Library (MCSTL)~\cite{singler:mcstl} that is now part of the GNU libstdc++ library. For the
base case of plane sweep algorithm, we use the C++ Standard Template Library (STL) sorting implementation.

\textbf{Choice of $P$.} While in theory, $P$ denotes the number of
cores, there are many considerations involved in picking the correct
value of the parameter $P$ in practise. These considerations are
discussed in Appendix~\ref{sec:choice_P}.

\section{Experiments}
\label{sec:experiments}
We performed extensive experimentation studying the performance of
these algorithms on various input types and on many different
multicore architectures. In addition to measuring the running time of
these algorithms, we used {\tt papi} library and the Linux perfctr kernel module to
read the hardware performance counters and measure cache misses, DRAM
accesses, TLB misses, branch mispredictions, number of instructions
etc. This section summarizes the key findings of
our experiments.

Our query points were generated uniformly at random inside the
grid of size $\mbox{Grid Size} \times \mbox{Grid Size}$. To elicit the
asymptotic worst case performance of point
location algorithms, we focus on long segments, whose length is chosen
uniformly at random between $\mbox{Grid Size/4}$ and $3\cdot\mbox{Grid
  Size}/4$ and are at a random $y$-coordinate.

\textbf{Configuration.} We ran our implementation on the following multicore systems:
\begin{enumerate}
\item
\label{item:intel}
A system with a single 4-core 2.66 GHz Intel Core i7-920 processor and a total
of 12.3GB RAM. Each core can run 2 threads due to hyperthreading. The
processor has an L3 cache of size 8192 KB that is shared
among all 4 cores. The L2 cache of 256 KB is only shared among pairs of cores.
\item
\label{item:amd}
A system with $4\times 12$-core 1.9 GHz AMD Opteron 6168 processors 
and total of 264 GB of RAM.  Each core contains a private L2 cache of 512
KB and groups of 6 cores share an L3 cache of 5118KB. Thus, each processor
contains two L3 caches of combined size of just over 10MB.
\item
\label{item:bli}
A system with 2 x 16-core 2.6 GHz AMD Opteron 6282 SE processors and
total of 96 GB RAM. Each core has its private L2 cache while the L3
cache is shared between 16 cores. The L2 cache size is 2 MB and L3 cache size is 16 MB. 
\end{enumerate}

All configurations run Linux kernels and the codebase was
compiled using g++-2.4 compiler and -O3 flag.

\textbf{Spatio-temporal locality in our setting}.
The cache line size for all cache levels on all 3 systems is 64 bytes. Since our objects take 32 bytes
of space, it appears that each cache line can hold only two objects.
Therefore, at a first glance it is not clear if I/O efficient
algorithm can utilize the spatial locality for any improvement in runtime. However, we
observed that given an array that is too large to fit in cache and which contains
our 32-byte objects, it takes 4-5 times faster to access the objects
sequentially rather than performing access in random locations. This
observation can be explained by the fact that the memory system prefetches 2-3
cache lines when performing a sequential scan. Thus, during sequential scan the
prefetcher amplifies the size of the cache line by the number of lines being
prefetched.\footnote{For this experiment, the array must contain the actual
objects and not just pointers to the objects, which could be allocated anywhere
in memory.}

Another benefit of performing $K$-way distribution sweeping is that it
allows us to utilize temporal locality by reducing the number of
recursive calls. In particular, $K$ is chosen as $K = \min \{n/M, M/B\}$ and
the number of recursive levels is $(1+\log_{K} (n/M))$. Given limit of RAM size
on our systems and the large size of L3 cache, it appears from our experiments
that $K$ is set to $n/M$ on configuration 1 and 2, resulting in a single recursive level
dedicated to (sequential) distribution (with the recursive base case performing
plane sweep on chunks that fit in L3 cache). On configuration 3, it
requires two recursive calls. The various trade-offs involved in
selecting the correct values of parameters $K$ and $M$ and the effect
of these parameters on the actual run-time of our PEM implementation
are described in Appendix~\ref{sec:param_km}.

\begin{figure*}[tb]
\center
\includegraphics[width=.3\textwidth, angle=270]{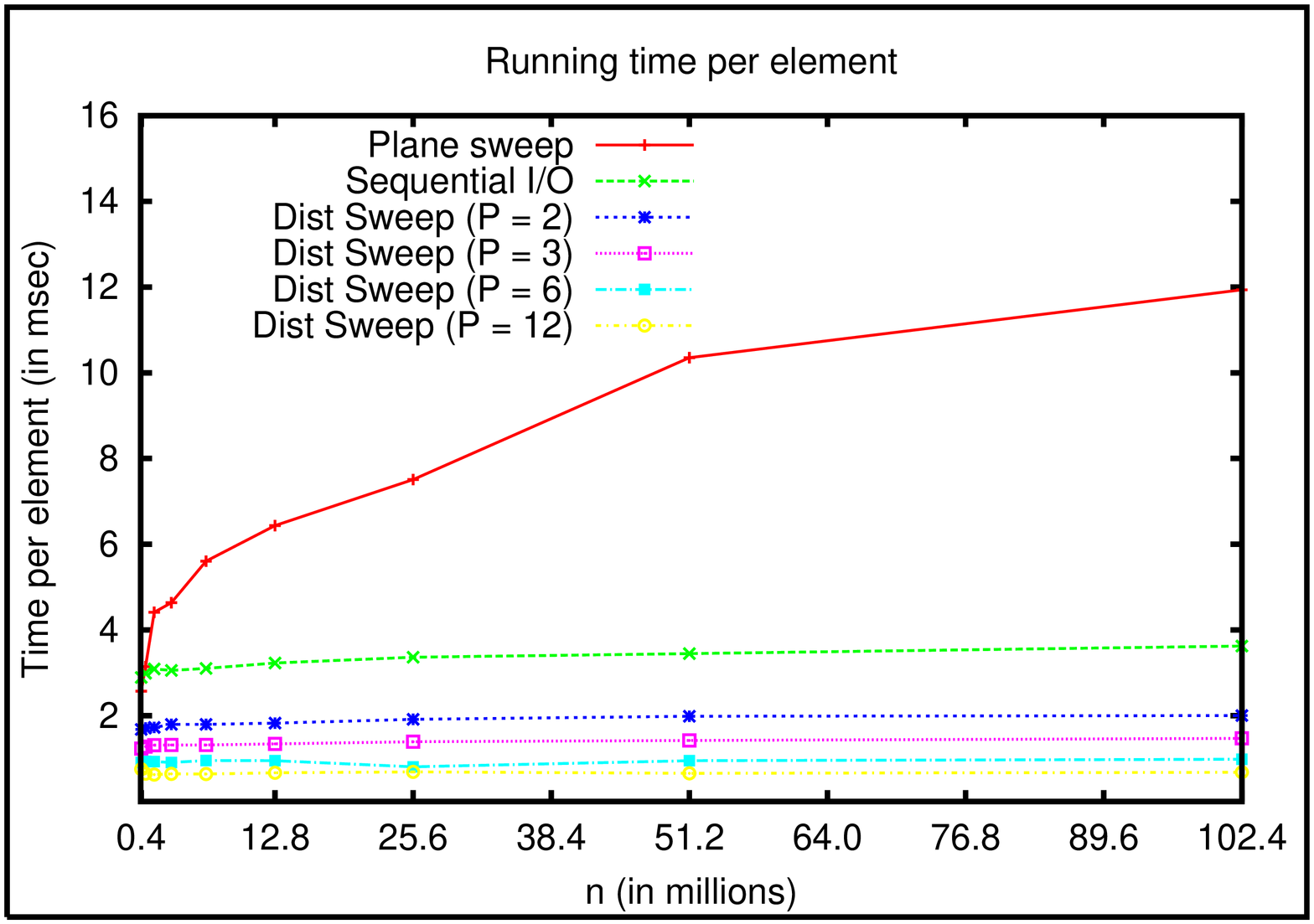}
\hspace{1cm}
\includegraphics[width=.3\textwidth, angle=270]{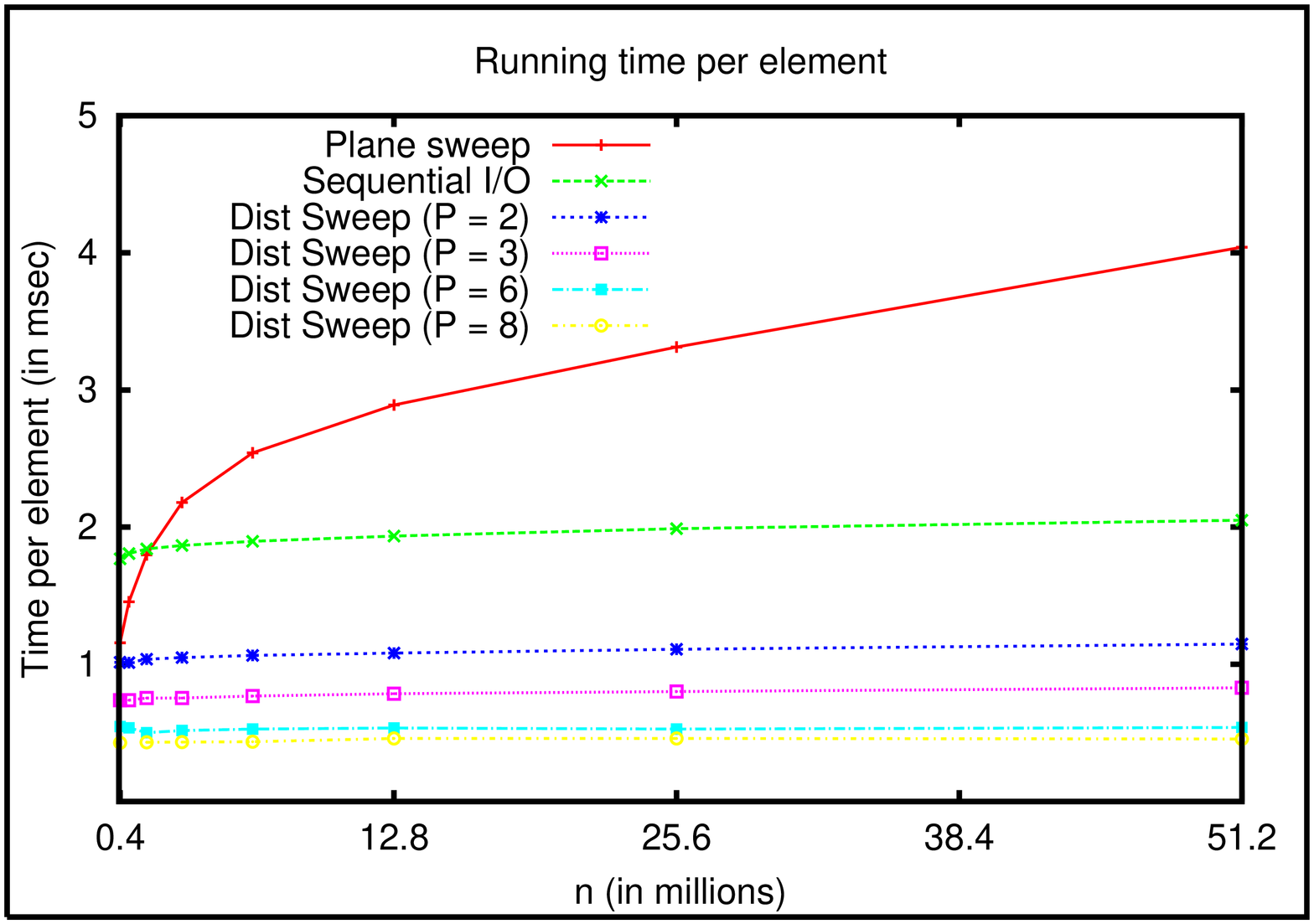}
\caption{Runtimes on the configuration~\ref{item:amd} (left) and
  configuration~\ref{item:intel} (right) per element. The plots exclude the times to perform initial
sorting of inputs by the $y$-coordinate for distribution sweeping and
$x$-coordinate for the plane sweep.}
\label{fig:runtimes}
\end{figure*}

\textbf{Random access vs. I/O-efficient algorithms.} 
 Figure~\ref{fig:runtimes} shows the absolute running times for the plane sweep
and (parallel) distribution sweeping algorithms. One can see improvements in
runtimes with the increase in the number of processors used. Also note the
difference in the slopes in the graphs of the plane sweep algorithm compared to
distribution sweeping algorithms. This is due to larger asymptotic number of
cache misses of the plane sweep algorithm. 

\begin{figure*}[!htb]
\center
\includegraphics[width=.3\textwidth, angle=270]{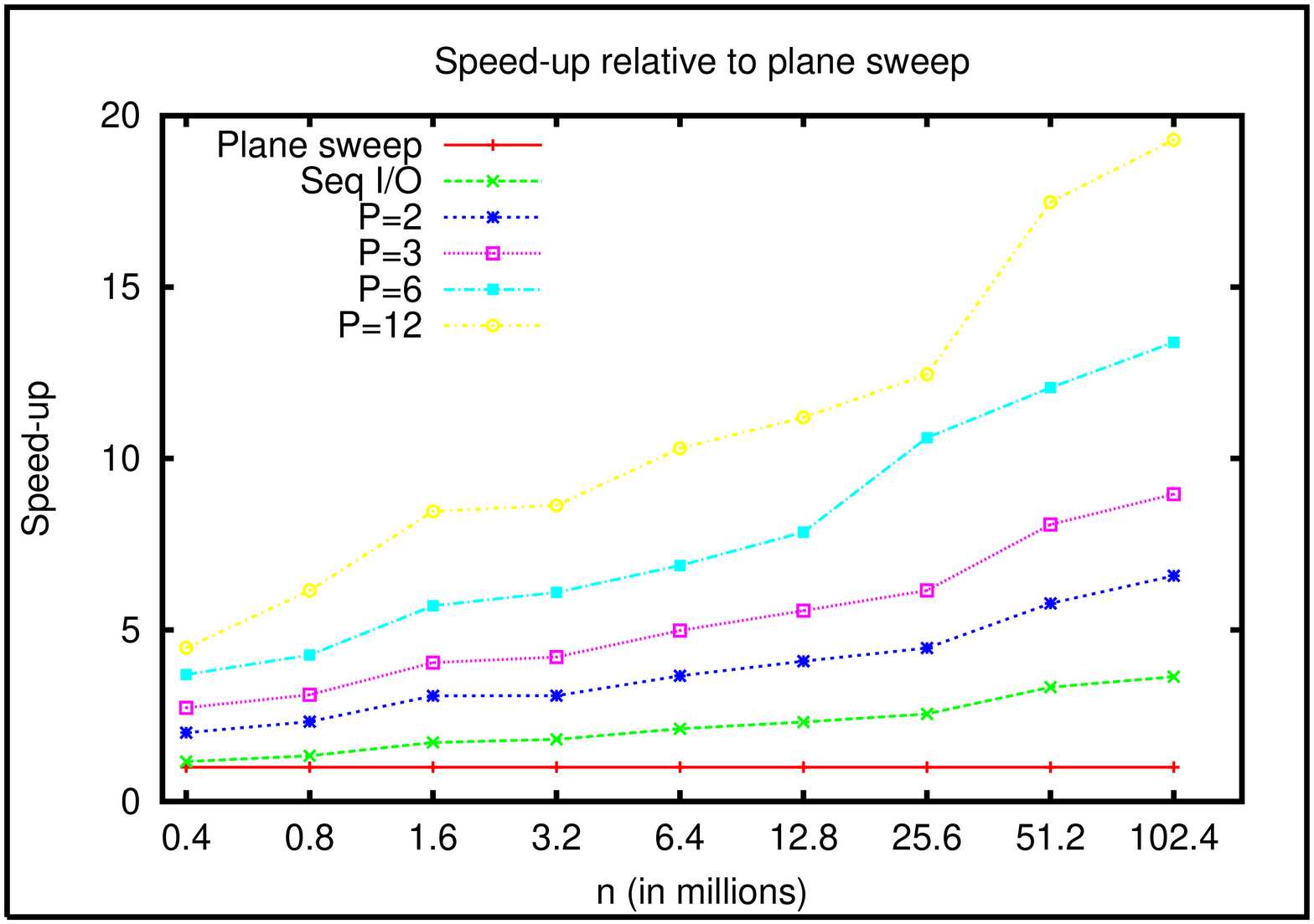}
\hspace{1cm}
\includegraphics[width=.3\textwidth, angle=270]{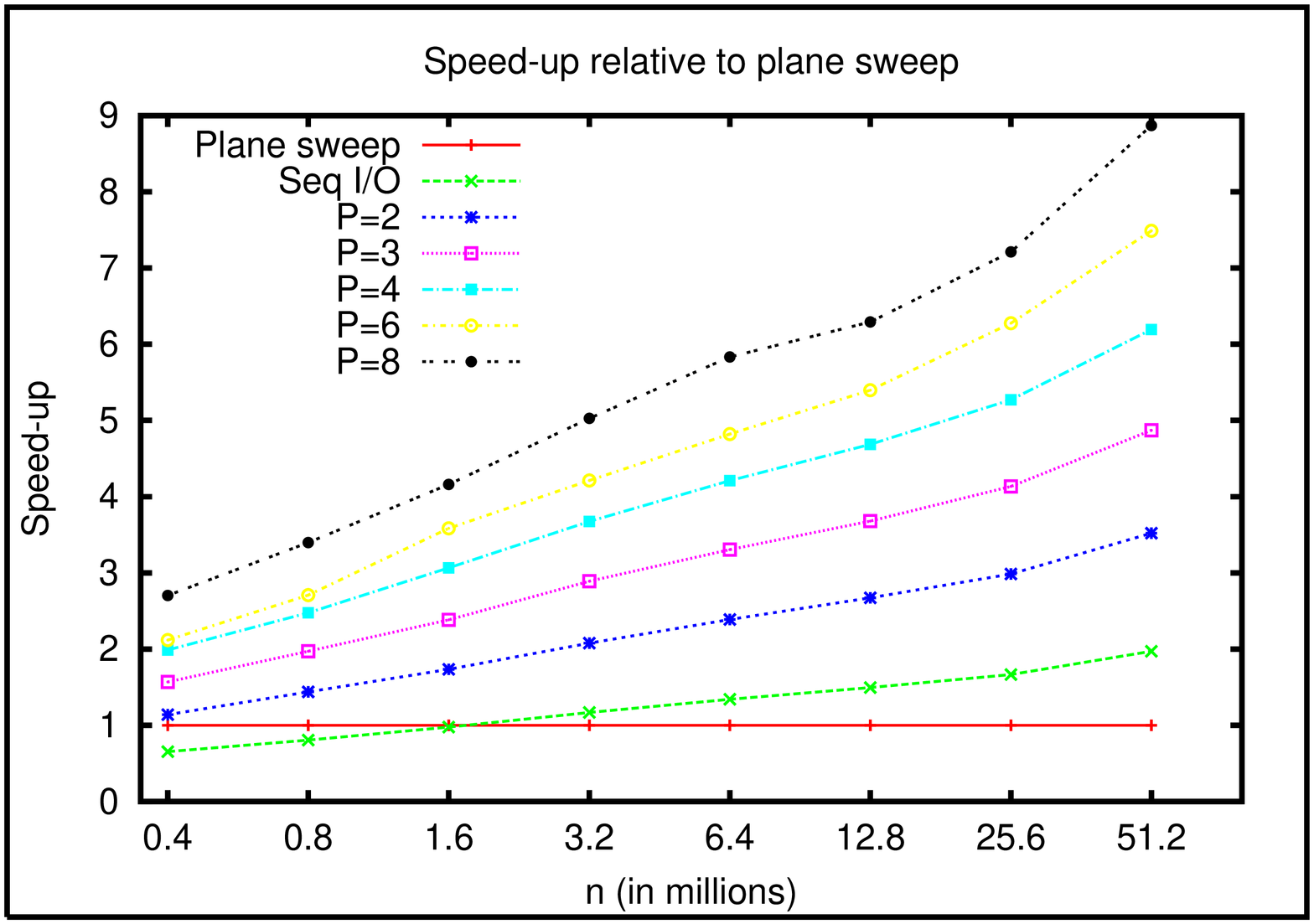}
\caption{Speedup of the distribution sweeping algorithms relative to the plane
sweep algorithm on the configuration~\ref{item:amd} (left) and configuration~\ref{item:intel} (right). The
plots exclude the times to perform initial sorting of inputs by the
$y$-coordinate for distribution sweeping and $x$-coordinate for the plane sweep.}
\label{fig:relative_linesweep}
\end{figure*}

Figure~\ref{fig:relative_linesweep} demonstrates this difference better. It
shows the speedup of the sequential and parallel distribution sweeping
algorithms relative to the plane sweep algorithm for long segments. In this
figure one can see the effects of cache-efficiency on runtimes.  It clearly
shows that the I/O-efficient algorithms outperform the plane sweep algorithm as
the input sizes increase. Recall our discussion that for the parameters of our
systems $K = n/M$ and the I/O complexity of the distribution sweeping algorithm
is $\OhOf{(n/B)(1+\log_K{n/M})} = \OhOf{n/B}$. This explains the non-linear
asymptotic speedup over plane sweep algorithm (with I/O complexity of
$\OhOf{(n/B) \log n/M}$) as a function of the input size.

\begin{figure*}[!htb]
\center
\includegraphics[width=.3\textwidth, angle=270]{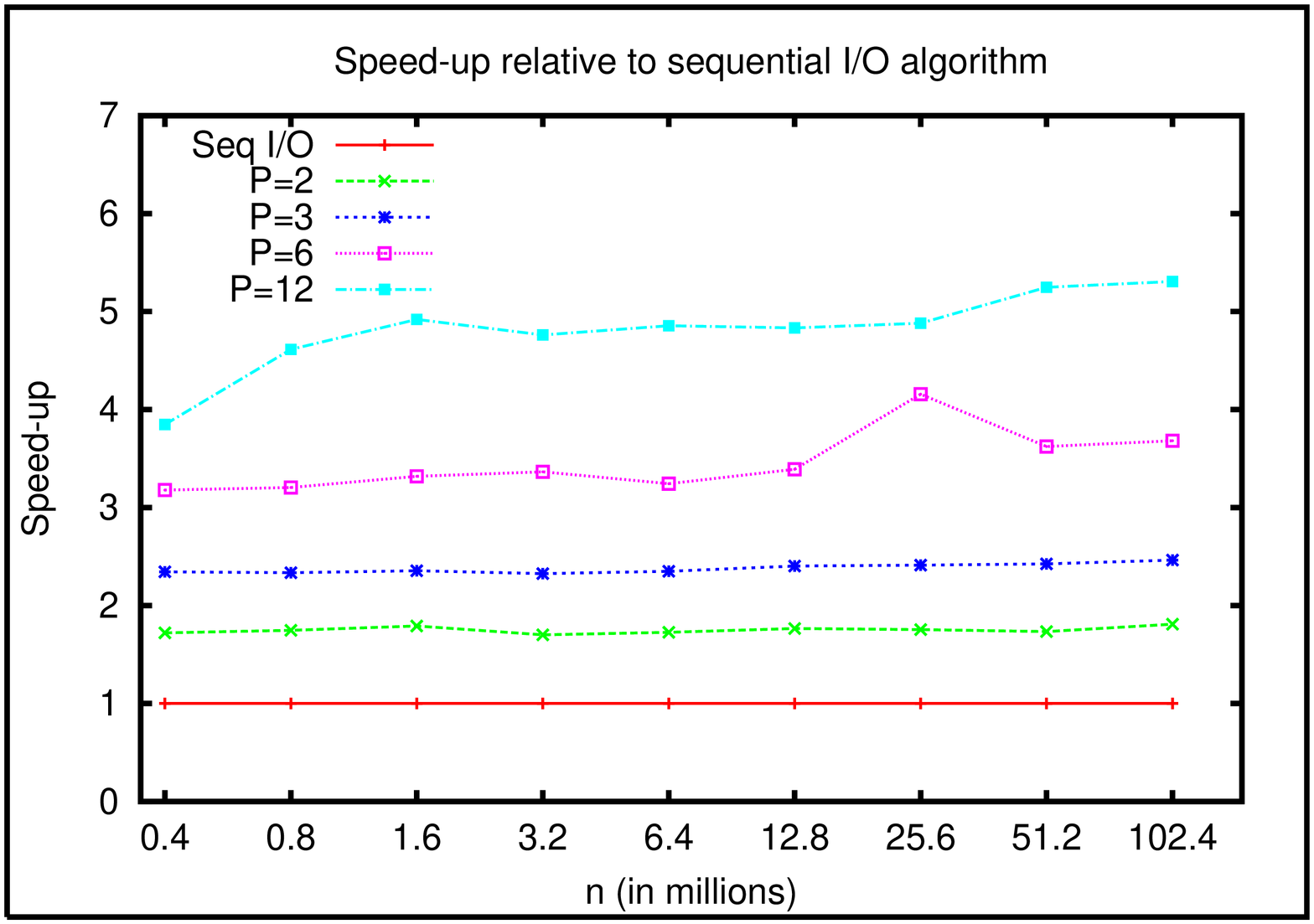}
\hspace{1cm}
\includegraphics[width=.3\textwidth, angle=270]{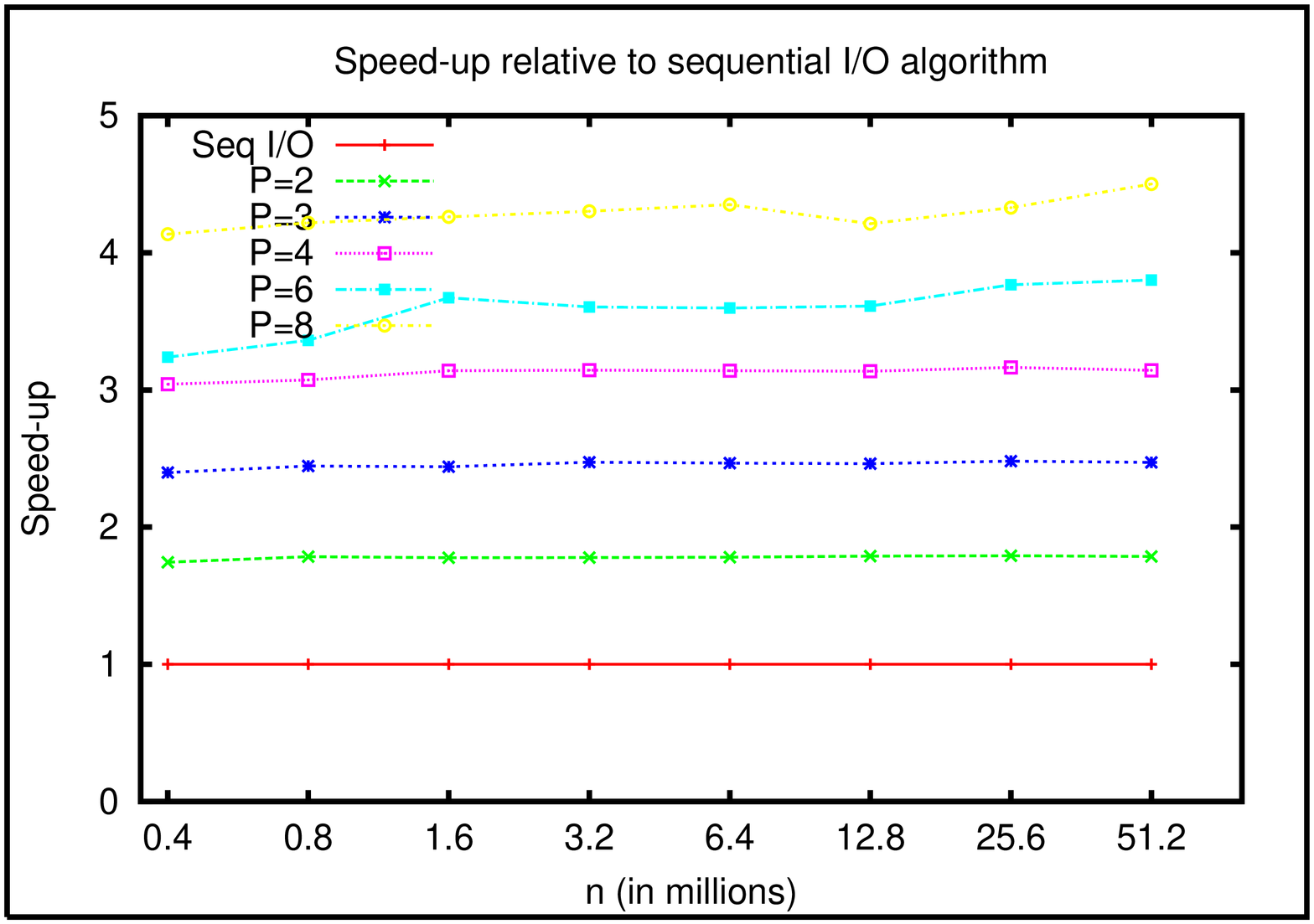}
\caption{Speedup of the parallel distribution sweeping algorithms relative to the sequential distribution sweeping
algorithm on configuration~\ref{item:amd} (left) and configuration~\ref{item:intel} (right) systems. The plots exclude the times to perform initial sorting of inputs by the $y$-coordinate.}
\label{fig:relative_sequential}
\end{figure*}

Figure~\ref{fig:relative_sequential} shows the speedup that parallel
distribution sweeping algorithm achieves relative to the sequential
distribution sweeping algorithm. 

\begin{figure*}[!htb]
\center
\includegraphics[width=0.3\textwidth, angle=270]{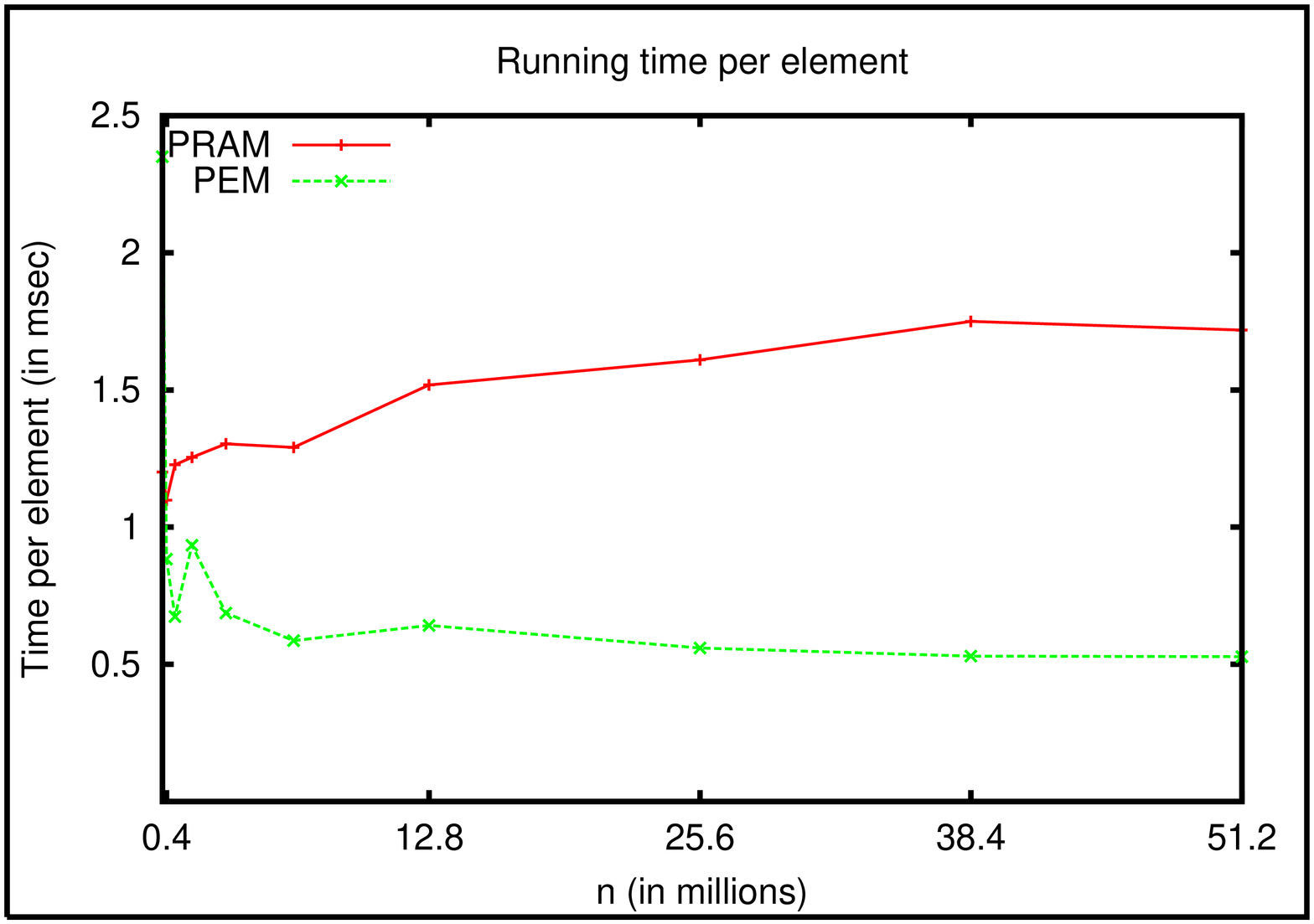}
\hspace{1cm}
\includegraphics[width=0.3\textwidth, angle=270]{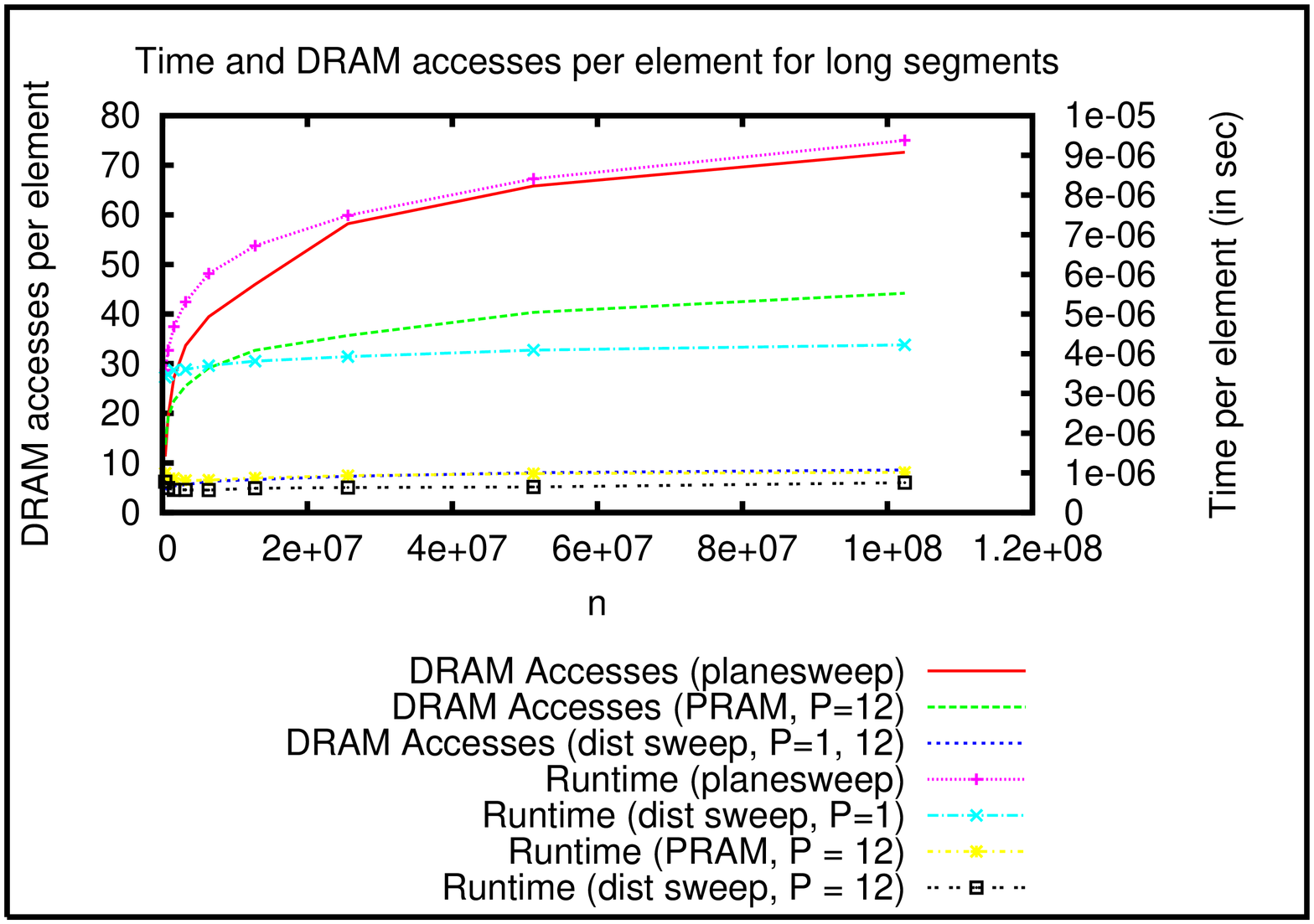}
\caption{Comparison of PEM and PRAM algorithms on 16 cores of
  configuration~\ref{item:bli} is shown in the left figure. Running time and DRAM
  traffic for long segments on 12 cores of configuration~\ref{item:amd}
  in the right.}
\label{fig:dram_access}
\end{figure*}

\textbf{PRAM vs. PEM performance.} Figure~\ref{fig:dram_access} (left) shows the comparative
performance of the various algorithms on configuration~\ref{item:bli}. We observe
that the PRAM implementation is significantly slower than the PEM
algorithm. For instance, with 51.2 million segments and the same
number of queries, PRAM implementation takes 96 seconds with 16
cores, while the PEM implementation only requires 30 seconds with the
same number of cores (excluding the time for loading the input and
sorting it, which is 18 seconds for both implementations). This is
largely accounted for by the fact that the PRAM implementation makes
poor use of temporal locality and thus, has larger number of recursive
levels. In each recursive level, it scans all the segments and query
points, increasing the DRAM accesses significantly. 

\textbf{DRAM Accesses and Cache Misses.} 
We could not find a reliable way to measure only L3 cache misses: the {\tt
papi} library does not support measurement of shared cache events, while the
hardware counters for LLC (Last Level Cache) counters returned suspiciously
similar results to L2 cache misses.  Instead we measured the total traffic to
DRAM using {\tt perf} tool.  Figure~\ref{fig:dram_access} (right) shows a clear
correlation between the total DRAM traffic and running times. It is interesting
to note that although our algorithms are designed in simple 2-level cache
model, they minimize the total traffic to DRAM, in spite of complex nature of
modern memory systems.

\textbf{Random, short, medium and long segments.}
We refer the reader to
Appendix~\ref{sec:long-short} for the relative behavior of the
different point location 
algorithms on different segment types.

\section{Conclusions And Future Work}
In this work, we explored the effects of caches on actual run-times
observed on various multicore
architectures in the context of the geometric stabbing-max query
problem. This is used to understand how accurately the PEM model predicts 
the running time of combinatorial algorithms on current multicore architectures.
On single-socket multicore architectures, our results show a direct
correlation between traffic on DRAM memory controller and running times of implementations. Thus, the algorithms designed
I/O-efficiently via the (parallel) distribution sweeping framework outperform
the plane sweep algorithms which do not address the I/O-efficiency. 

We chose to perfom our experiments on single-socket architectures, because the
PEM model assumes uniform access latencies to shared memory.
We conjecture that NUMA effects of DRAM access on multi-socket architectures
might be better modeled by distributed computational models, where
each processor copies/moves data into ``local'' memory --- address space
associated with its socket --- before processing it. Once the data is in its
``local'' memory, one can use the PEM model to design algorithms to process the
data. The experimental evaluation and modeling NUMA effects of multi-socket
architectures is left for future investigations.

While we chose to implement an algorithm which was designed in the PEM model,
it would be interesting to see how the implementations in other cache-conscious
parallel models (for example,~\cite{blelloch:multicore-dnc}) will fare in
practice in similar setting. 

\textbf{Acknowledgments.} We would like to thank Peter Sanders for encouraging
to look at the work-optimality of PEM algorithms. We would also like to thank  Dennis
Luxen and Dennis Schieferdecker for their extensive help with our
implementations and getting {\tt perf} and {\tt papi} to run on our systems.

\bibliographystyle{abbrv}
\bibliography{refs}

\pagebreak 

\begin{appendix}

\section{Details of the PEM solution} \label{app:pem-solution}

In this section we present the details of the PEM solution. 

As mentioned earlier, our algorithm is based on the {\em parallel} distribution
sweeping framework introduced by Ajwani et al.~\cite{ajwani:pem-geom}. Parallel
distribution sweeping recursively divides the plane into vertical slabs,
starting with the entire plane as one slab and in each recursive step dividing
a given slab into $K := \max\{2, \min\{\sqrt{n/P}, {M/B}, P\}\}$ child slabs.
This division is chosen so that each slab at a given level of recursion
contains roughly the same number of objects (segment endpoints and query
points).  The first level of recursion divides the plane into $P$ slabs, each
containing $\ThetaOf{n/P}$ input elements.  Viewing the recursion as a rooted
tree defines leaf invocations and children of a non-leaf invocation.  An
invocation on slab $\sigma$ at the $k^{th}$ recursive level is denoted as
$I^k_\sigma$. 

Each invocation $I^k_\sigma$ receives as input a $y$-sorted list $Y^k_\sigma$
containing segments and query points. The root
invocation $I^0_{\reals^2}$ contains all segments and query
points of the input and the input $Y^0_{\reals^2}$ is generated by sorting the
horizontal segments and query points by the $y$-coordinate. For a non-leaf invocation $I^k_\sigma$, let
$I^{k+1}_{\sigma_1}, I^{k+1}_{\sigma_2}, \dots, I^{k+1}_{\sigma_K}$ denote its
child invocations. The input $Y^{k+1}_{\sigma_j}$ for a child
invocation $I^{k+1}_{\sigma_j}$ consists of the
$y$-sorted list of segments in $Y^k_\sigma$ with an endpoint in
$\sigma_j$ and the query points in $\sigma_j$. 
In processing $I^k_\sigma$, we consider all the 
children slabs of $\sigma$: $\sigma_1, \sigma_2, \dots,
\sigma_K$ and compute a segment
$l^q_{\sigma_j} \in Y^k_\sigma$ for all query points $q \in \sigma_j$, that is the highest segment lower than $q$ and spans
$\sigma_j$. If $l^q_{\sigma_j}$ is higher than $r_q$, we let $r_q :=
l^q_{\sigma_j}$. Thus, $r_q$ always stores the highest
segment lower than $q$ that spans $\sigma_j$ (and all its parent slabs).

At every leaf invocation $I^k_{\sigma}$, the highest segment below the
query points are found using sequential I/O-efficient distribution
sweeping technique as described in the previous section. This is then
compared to the current stored value $r_q$ and the maximum of the two values is stored in $r_q$.

To compute the values of $l^q_{\sigma_i}$ for all child slabs, we
process them using the $P$ processors as follows: We partition the
input sets $Y = \cup_i Y^k_{\sigma_i}$ into $P$ equal chunks $Y_1,
Y_2, \dots, Y_P$ based on the $y$-coordinate, each one of size
$\ThetaOf{n/P}$. Then, processor $p_j$ processes
    $Y_j$ using the sequential I/O-efficient algorithm independently of others.  Note, that this
    process sets the initial values of $s_{\sigma_i}$ to $l_{-\infty}$ -- a virtual horizontal line $y=-\infty$. Thus, if at the
    end of the process $l^q_{\sigma_i} = l_{-\infty}$ for some query $q \in Y_j$ and there is a
    segment $s \in Y_k$, $k < j$ such that $s$ lies below $q$, then $l^q_{\sigma_i}$ is still not set correctly. To fix
    this, at the end of sequential sweep by each processor $p_j$, $p_j$ saves
    the values of $s_{\sigma_i}$ for each of its slabs. These values are then
    processed by all processors in a way similar to segmented prefix sums to
    propagate the values of $s_{\sigma_i}$ to the appropriate processors as
    follows. Assume, the segment identifiers of segments increase with the
    increase in segments' $y$-coordinates. Now consider $K$ independent prefix
    sums with $\max$ as the associative operator applied on the $K \cdot P$
    values of $s_{\sigma_i}$ (one prefix sums on the values within a single
	slab).  Finally, we initialize the value of $s_{\sigma_i}$ at each
    processor $p_j$ to the final value (after the prefix sums) of $s_{\sigma_i}$
    at processor $p_{j-1}$ (in case of $p_0$, $s_{\sigma_i} = l_{-\infty}$ for all
	$\sigma_i$) and repeat the sequential sweep by each processor. Note, the
    purpose of the prefix sums is to propagate the values of last seen segment
    across multiple processors and after the second sweep, all $l^q_{\sigma_i}$ are set
    correctly.

The parallel I/O complexity of the above algorithm is
$\OhOf{\psort{n}}$ I/Os.

\section{Data representation and space efficiency of our
  implementations}
\label{sec:space}
We implement the segments and queries as a single vector of objects.  To
achieve this, we represent each line segment and query point with a single
32-byte structure as follows. Using double precision for the coordinates, we
need 16 bytes to represent a point on a plane. Additional 8 bytes are used for
each segment to represent the $x$-coordinate of the second endpoint or for each
query to record the $y$-coordinate of the segment below it as an answer.
Additional 4 bytes are used for the identifier of each segment (we do not
assume that segments have unique $y$ coordinates, and therefore, they cannot be
identified with this ID field).  The same field entry is used for recording the
segment ID of the output for a query. Finally, we need at least one boolean
value for distinguishing a segment from a query. However, since memory
allocation in C++ is aligned at 4 byte memory intervals, the last field of the
structure takes up at least 4 bytes.  While keeping segments and points as a
single object type might cost extra 4 bytes of memory for each item, this
approach has the advantage of simplifying computations and internal data
structures for the slabs as we no longer need to keep separate lists for query
points and line-segments, but can keep a single sorted list. 

Let $s$ be the number of segments in the input and $q$ be the number of query
points in the input, i.e., $n = s+q$. The sequential distribution sweeping can be implemented to
use space taken up by up to $3s+2q$ objects. This number arises from the fact
that during distribution we might create up to two copies of each segment (one
for each end point) to be placed into child slabs. In addition the distribution
at each recursive level cannot be performed in place and, hence, we need to
allocate additional $2s+q$ memory during the process. In practice to achieve
this bound, one needs to know exactly how many objects will be distributed to
each child slab. This would require an additional pass over the data to count
the the sizes of each child slabs, increasing the running time by a factor of
2.  Instead, we use dynamic arrays (e.g., vectors in C++ STL) which grow
automatically when the data exceeds the preallocated capacity. During the
resizing, the contents of vectors are copied over into the newly allocated
vector, seemingly resulting in the same double the running time. However,
copying a vector is performed using low level memory copying routines which are
more efficient than traversing the input twice. In addition, we utilize the
fact that we run our experiments on uniformly distributed data. By
allocating 10\% more space than 1/K-th fraction of the total data to be
distributed to $K$ children, with high probability no child exceeds the
preallocated space. 

\section{On the right choice of $P$} \label{sec:choice_P} 
The PEM model prescribes algorithms for an architecture which contains a
private cache per processor connected with an independent channel to the shared
memory. The PEM algorithms measure the number of parallel I/Os performed.  If
the bandwidth of channels is fully utilized, increasing the number of
processors without increasing the number of memory channels would not result in
reduced parallel I/Os. In reality, modern multicores have much fewer memory
channels than there are cores so an interesting question is whether it is
beneficial to increase the number of processors beyond the number of memory
channels.  

The Intel i7 system contains 3 memory channels and implements hardware counters
which record overall DRAM accesses per each channel. Interestingly, throughout
the computation, one of the channels recorded no DRAM accesses, while the other
two channels shared the traffic to DRAM unevenly. The discrepancy in traffic
load between the two channels decreased with higher number of cores used. We cannot explain the reason why the system did not use all memory channels and this is worth further investigations.

We could not measure the bandwidth utilization of the memory channels. However,
we did measure the IPC -- an average number of instructions executed per clock
cycle. The results showed IPC on the Intel i7 system being close to 1
regardless of number of cores used for the distribution sweeping
implementation. However, for the planesweep implementation the IPC dropped down
to .7 for large inputs. This leads us to believe that the random access of
planesweep results in inefficient use of the memory bandwidth.  At the same
time full cache line transfers of the distribution sweeping implementation
barely saturates the memory channels even with all 4 cores running 2 threads
each.

\section{Effect of parameters $K$ and $M$ on runtime.} 
\label{sec:param_km}
\begin{figure*}
\center
\includegraphics[width=.3\textwidth, angle=270]{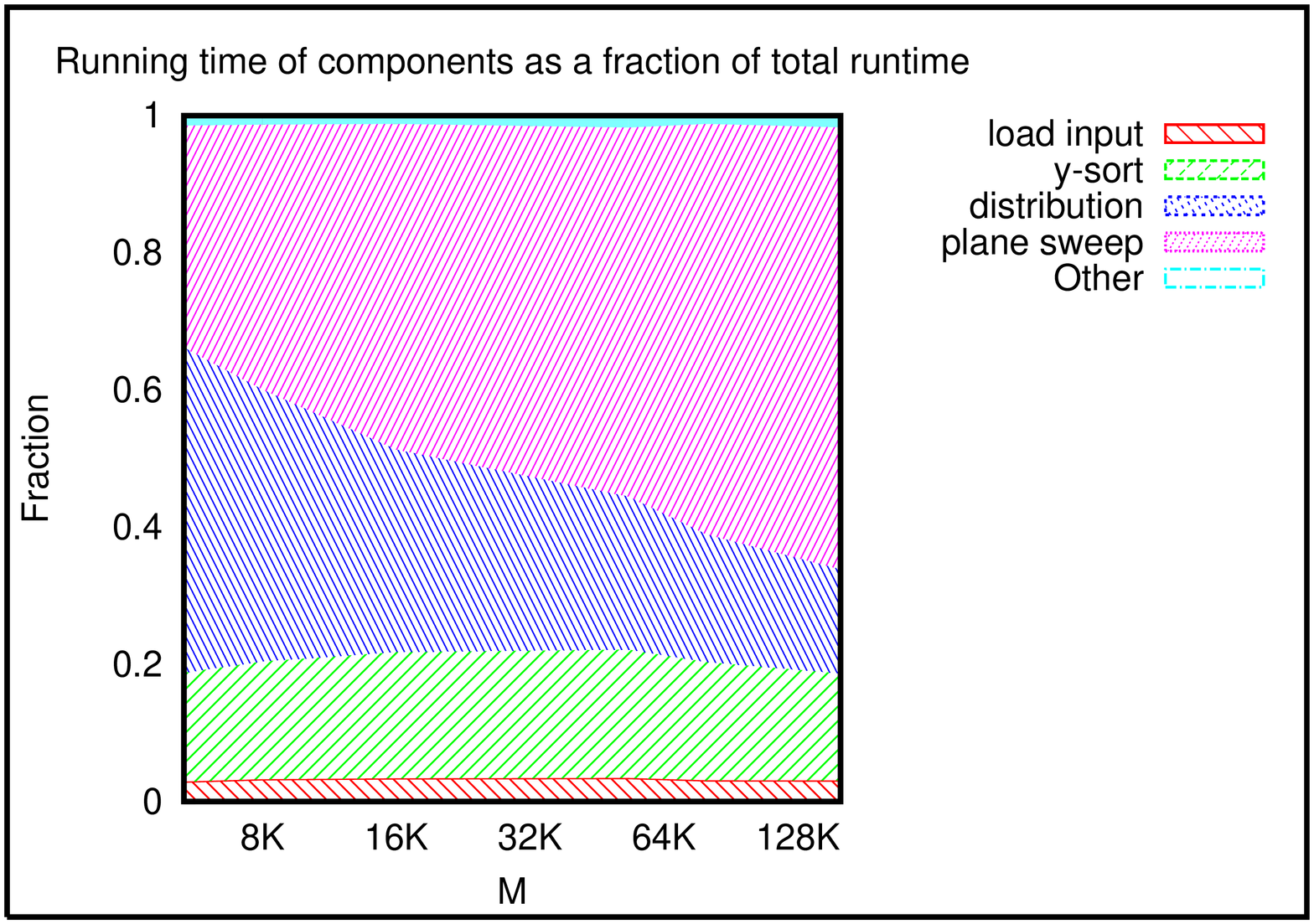}
\hspace{1cm}
\includegraphics[width=.3\textwidth, angle=270]{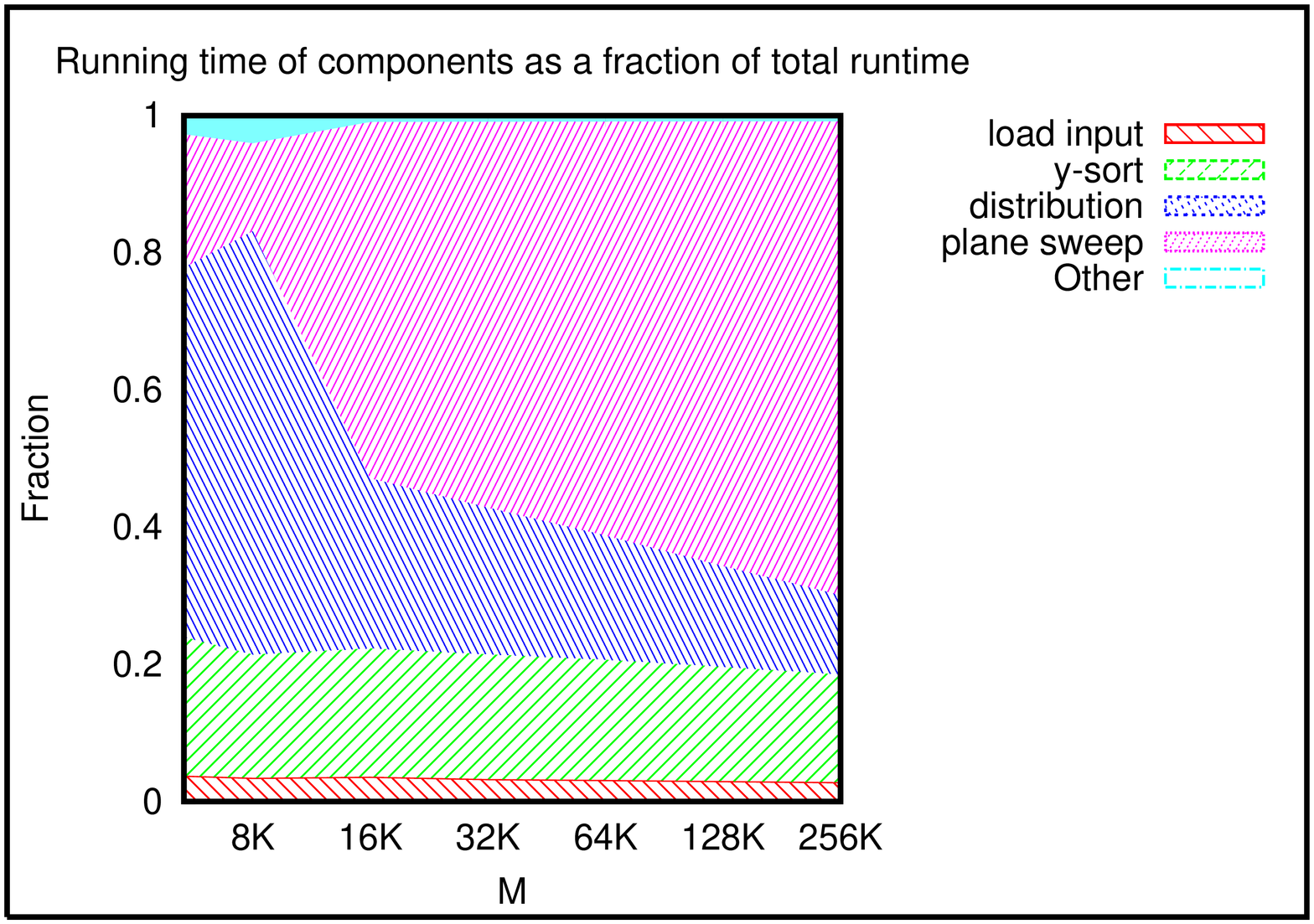}
\caption{Task break-down as a percentage of the total running time for a sequential distribution sweeping 
on the AMD Opteron 6168 with $n = 102.4\cdot 10^6$ objects (left) and Intel i7 with $n = 51.2\cdot 10^6$ objects (right). The different tasks are (a) loading the input from file, (b) initial sorting of the input by $y$-coordinate, (c) total time taken by distribution of of objects into child slabs, (d) total time take by plane sweep algorithm at the base of the recursive calls and (e) misc. bookkeeping not included in all of the above.}
\label{fig:relative_work}
\end{figure*}
Figure~\ref{fig:relative_work} shows how much each task takes as a fraction of
total runtime as we vary the threshold parameter $M$.  One can see that by
reducing $M$ down to the L2 cache size results in faster combined execution
time of all invocations of the plane sweep algorithm at the base case of the
recursion. This is due to the fact that the binary search tree $T$ used for
the plane sweep algorithm fits in the faster L2 cache. 

However, this decrease in runtime of the base case is offset by the increase in
the runtime of the distribution sweeping phase due to the following reason.
Recall that $K = \min\{n/M, M/B\}$. If $K = n/M$, larger value of $M$ results
in increase of $K$ -- the number of slabs to distribute the objects at each
recursive level. While the number of slabs is still small enough to fit in
cache, and, therefore, there is no increase in cache misses, the time it takes
to identify the slab where to distribute each item takes $\OhOf{\log K}$
internal computation time, i.e. it grows with $K$.  If, on the other hand, $M$
decreases so much that $K = M/B$, the number of recursive levels grows as a
function of $\log_K (n/M)$. Thus, decrease in $K$ results in more recursive
levels which in turn results in more scans of the input and, therefore, more
(capacity) cache misses. 

We also observed that setting $M$ equal to exactly the size of a cache does not
result in the best runtimes. This can be explained by the more complex nature
of caches, such as set associativity and the replacement policy: the external
memory model assumes fully associative cache with optimal replacement policy,
while modern architectures implement set-associative caches with (most
likely)\footnote{It is hard to determine the true replacement policy because processor
manufacturers keep this information confidential.} the Least Recently Used
(LRU) replacement policy. To achieve the best results we set $M$ to a quarter
of the L3 cache size for the Intel i7 architecture and a third of the L3 cache
size for the AMD Opteron architectures. 

Since the caches are shared among subsets of processors, in case of the
parallel execution, our initial intuition was to reduce $M$ by the number of
processors sharing the cache. However, our experiments showed that this is
unnecessary and the same $M$ as for the sequential implementation works just as
well in the parallel implementation. 

This can be explained by the fact that in our experiments, $K = n/M$ which
constitutes a much smaller portion than the L3 cache. Therefore, during the
parallel distribution sweeping, maintaining one block for each
child slab in cache and maintaining for each processor the tree $T'$ of size $2\cdot K$ for
work-optimal distribution sweeping does not interfere with other processors'
cache data.

\section{Effect of segment type on the performance of point location algorithms}
\label{sec:long-short}
To analyze the behavior of our algorithms on segments of varying
lengths, we first generate the line-segments in different ways:
 
\textbf{Random segments: }
Our first input set is a set of random line-segments in the 
grid. The random lines are generated by selecting a random
y-coordinate and two random x-coordinates in the grid. Thus, the expected
length of the line segments is $\OhOf{(\mbox{Grid
    Size})}$.

\textbf{Short segments:}
Here, $n$ line segments are generated with length chosen uniformly at random
between $\mbox{Grid Size}/n$ and $4\cdot\mbox{Grid Size}/n$.

\textbf{Medium segments:} 
Here, $n$ line segments are generated with length chosen uniformly at random between
$\mbox{Grid Size}/\sqrt{n}$ and $4\cdot\mbox{Grid Size}/\sqrt{n}$.

\textbf{Long segments:}
Here, we generate line segments with lengths chosen uniformly at random between
$\mbox{Grid Size/4}$ and $3\cdot\mbox{Grid  Size}/4$. 

\begin{figure*}[!ht]
\center
\includegraphics[width=.3\textwidth, angle=270]{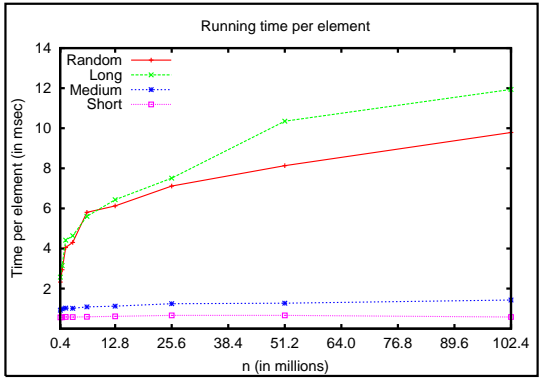}
\hspace{1cm}
\includegraphics[width=.3\textwidth, angle=270]{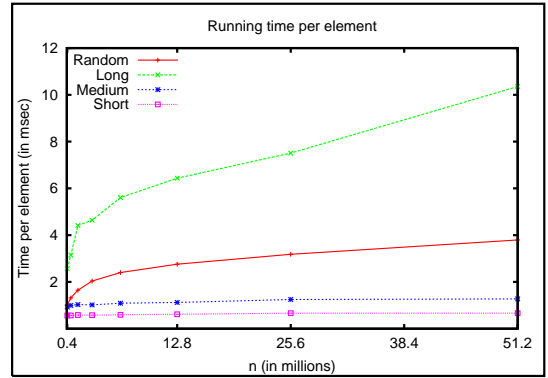}
\caption{Runtimes for small, medium, long and randomly sized segments for the plane sweep on AMD Opteron 6168 (left) and Intel i7 (right) systems as a function of input size.}
\label{fig:segment_sizes_planesweep}
\end{figure*}

\begin{figure*}
\center
\includegraphics[width=.3\textwidth, angle=270]{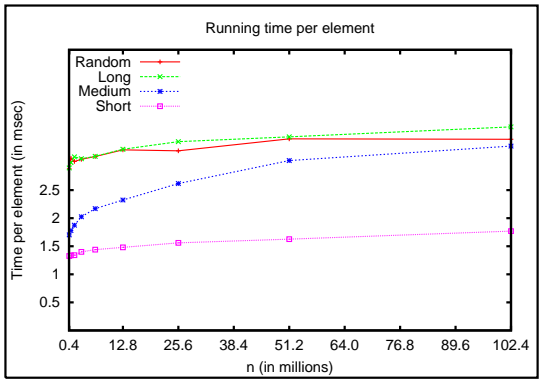}
\hspace{1cm}
\includegraphics[width=.3\textwidth, angle=270]{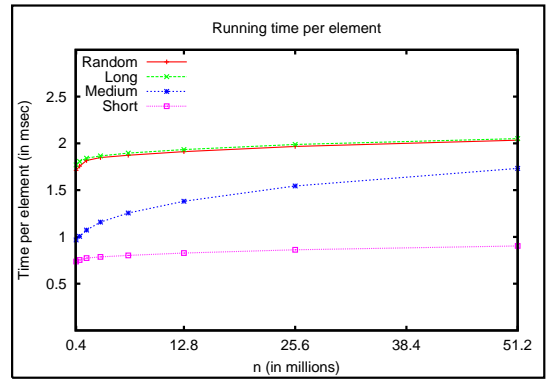}
\caption{Runtimes for small, medium, long and randomly sized segments for the sequential distribution sweeping on AMD Opteron 6168 (left) and Intel i7 (right) systems as a function of input size.}
\label{fig:segment_sizes_sequential_io}
\end{figure*}
\begin{figure*}[!ht]
\center
\includegraphics[width=.3\textwidth, angle=270]{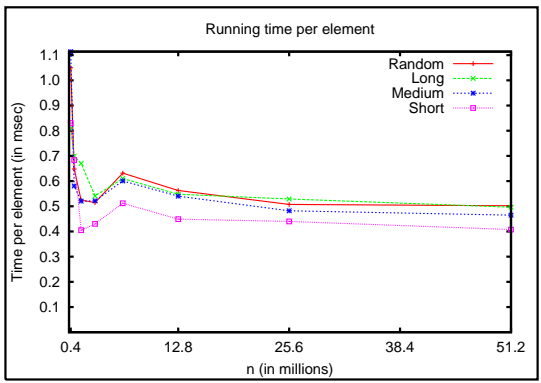}
\hspace{1cm}
\includegraphics[width=.3\textwidth, angle=270]{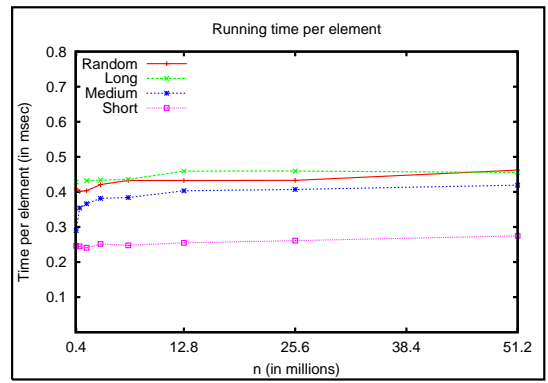}
\caption{Runtimes for small, medium, long and randomly sized segments for the
parallel distribution sweeping on AMD Opteron 6168 (left) and Intel i7 (right)
systems as a function of input size. The results are for the maximum number of threads for each system.
}
\label{fig:segment_sizes_parallel_io}
\end{figure*}

In Figures~\ref{fig:segment_sizes_planesweep} through
~\ref{fig:segment_sizes_parallel_io} one can see how different sizes of
segments affect the running times of our different implementations. We observe
that the plane sweep (Figure~\ref{fig:segment_sizes_planesweep}) performs much
worse on the long segments than on the short segments. This is because the
expected number of short segments intersecting any vertical line is expected to
be constant and the set of {\em active segments} $A$ fits in cache at all times. Thus, the
updates and predecessor queries on $A$ do not incur any additional cache misses. On the
other hand, the number of long segments intersecting any vertical line is
expected to be linear with the input size and the traversal of $T$ will incur
a lot of cache misses. Contrast this with the runtimes for sequential
(Figure~\ref{fig:segment_sizes_sequential_io}) and 
parallel
(Figure~\ref{fig:segment_sizes_parallel_io}) distribution sweeping algorithms
for the different segment sizes, which show much smaller variance of runtimes
as a function of different segment lengths. This confirms that the results that
Chiang~\cite{chiang:exp_em,chiang:exp_em2} observed 17 years ago still hold on
modern architectures.

\end{appendix}

\end{document}